\documentclass[10pt,a4paper]{article}

\usepackage[margin=2.5cm]{geometry}
\usepackage[skip=6pt,parfill]{parskip}

\usepackage{mathptmx}
\usepackage[T1]{fontenc}
\usepackage[utf8]{inputenc}
\usepackage{textcomp}
\usepackage{microtype}

\usepackage{amsmath,amssymb,amsfonts,mathtools}
\numberwithin{equation}{section}

\usepackage{graphicx}
\usepackage{subcaption}

\usepackage{float}
\usepackage[font=footnotesize,labelfont=bf,labelsep=colon]{caption}
\usepackage{tabularx}
\usepackage{booktabs}
\AtBeginEnvironment{tabular}{\small}
\AtBeginEnvironment{tabularx}{\small}

\usepackage[hidelinks]{hyperref}
\usepackage{xurl}
\usepackage[
  backend=biber,
  style=numeric,
  citestyle=numeric-comp,
  sorting=none,
  giveninits=true,
  maxnames=6,
  minnames=1,
  doi=true,  
  url=false,
  eprint=false,
  isbn=false
]{biblatex}
\AtEveryBibitem{%
  \clearlist{language}%
  \clearfield{issn}%
  \clearfield{isbn}%
  \clearfield{url}%
  \clearfield{eprint}%
  \clearfield{note}%
  \clearfield{addendum}%
  \clearfield{publisher}%
  \clearfield{address}%
  \clearfield{urldate}%
  \clearfield{month}%
  \clearfield{issue}%
  \clearfield{number}%
}

\DeclareFieldFormat[article]{journaltitle}{\mkbibemph{#1}}
\DeclareFieldFormat[article]{volume}{\mkbibbold{#1}}
\DeclareFieldFormat{pages}{#1}
\DeclareFieldFormat{labelnumberwidth}{#1.}
\DeclareFieldFormat{doi}{\href{https://doi.org/#1}{#1}}

\renewbibmacro*{begentry}{%
  \ifkeyword{special}{\printtext{\textbf{* }}}{}%
  \ifkeyword{outstanding}{\printtext{\textbf{** }}}{}%
}

\renewbibmacro*{finentry}{%
  \setunit{\par\nobreak\vspace{2pt}}%
  {\small\printfield{annotation}}%
  \finentry
}

\DeclareBibliographyDriver{article}{%
  \usebibmacro{begentry}%
  \printnames{author}%
  \setunit{\labelnamepunct}\newblock
  \printfield[titlecase]{title}%
  \setunit{\adddot\space}%
  \printfield{journaltitle}%
  \setunit{\addspace}%
  \printfield{year}%
  \setunit{\addcomma\space}%
  \printfield[volume]{volume}%
  \setunit{\addcolon}%
  \printfield{pages}%
  \setunit{\adddot\space}%
  \printfield{doi}%
  \usebibmacro{finentry}%
}

\usepackage[bottom,hang,flushmargin]{footmisc}

\title{\textbf{Design of DNA Strand Displacement Reactions}}

\author{%
  Križan Jurinović\footnotemark[1] \and
  Merry Mitra\footnotemark[1] \and
  Rakesh Mukherjee\footnotemark[1] \and
  Thomas E.~Ouldridge\footnotemark[1]\kern0.5em\footnotemark[2]
}

\date{}

\begin{document}

% ------------------------------------------------------------
% Include separate title page
% ------------------------------------------------------------
% ------------------------------------------------------------
% Title page
% ------------------------------------------------------------

\begingroup
\renewcommand{\thefootnote}{\fnsymbol{footnote}}
\renewcommand{\footnoterule}{\kern -3pt \hrule width \textwidth \kern 2.6pt}
\maketitle

% Extra space between authors and abstract
\vspace{1.5em}

% Abstract block without indentation
\begin{abstract}
\noindent
DNA strand displacement (SD) reactions are central to the operation of many synthetic nucleic acid systems, including molecular circuits, sensors, and machines. Over the years, a broad set of design frameworks has emerged to accommodate various functional goals, initial configurations, and environmental conditions. Nevertheless, key challenges persist, particularly in reliably predicting reaction kinetics. This review examines recent approaches to SD reaction design, with emphasis on the properties of single reactions, including kinetics, structural factors, and limitations in current modelling practices. We identify promising innovations while analysing the factors that continue to hinder predictive accuracy. We conclude by outlining future directions for achieving more robust and programmable behaviour in DNA-based systems.
\end{abstract}

% Extra space between abstract and keywords
\vspace{1.5em}

\begin{center}
\end{center}

\footnotetext[1]{Department of Bioengineering and Imperial College Centre for Engineering Biology, Imperial College London, Exhibition Road, London SW7 2AZ, UK.}
\footnotetext[2]{Corresponding author: Thomas E.~Ouldridge, \texttt{t.ouldridge@imperial.ac.uk}}

\endgroup

% Reset numbering for main text
\setcounter{footnote}{0}
\renewcommand{\thefootnote}{\arabic{footnote}}

% ------------------------------------------------------------
% Main text
% ------------------------------------------------------------
\newpage

\section{Introduction}

\begin{figure}[htbp]
    \centering
    \setlength{\abovecaptionskip}{4pt plus 1pt minus 1pt}
    \setlength{\belowcaptionskip}{2pt plus 1pt minus 1pt}
    \includegraphics[width=0.75\textwidth]{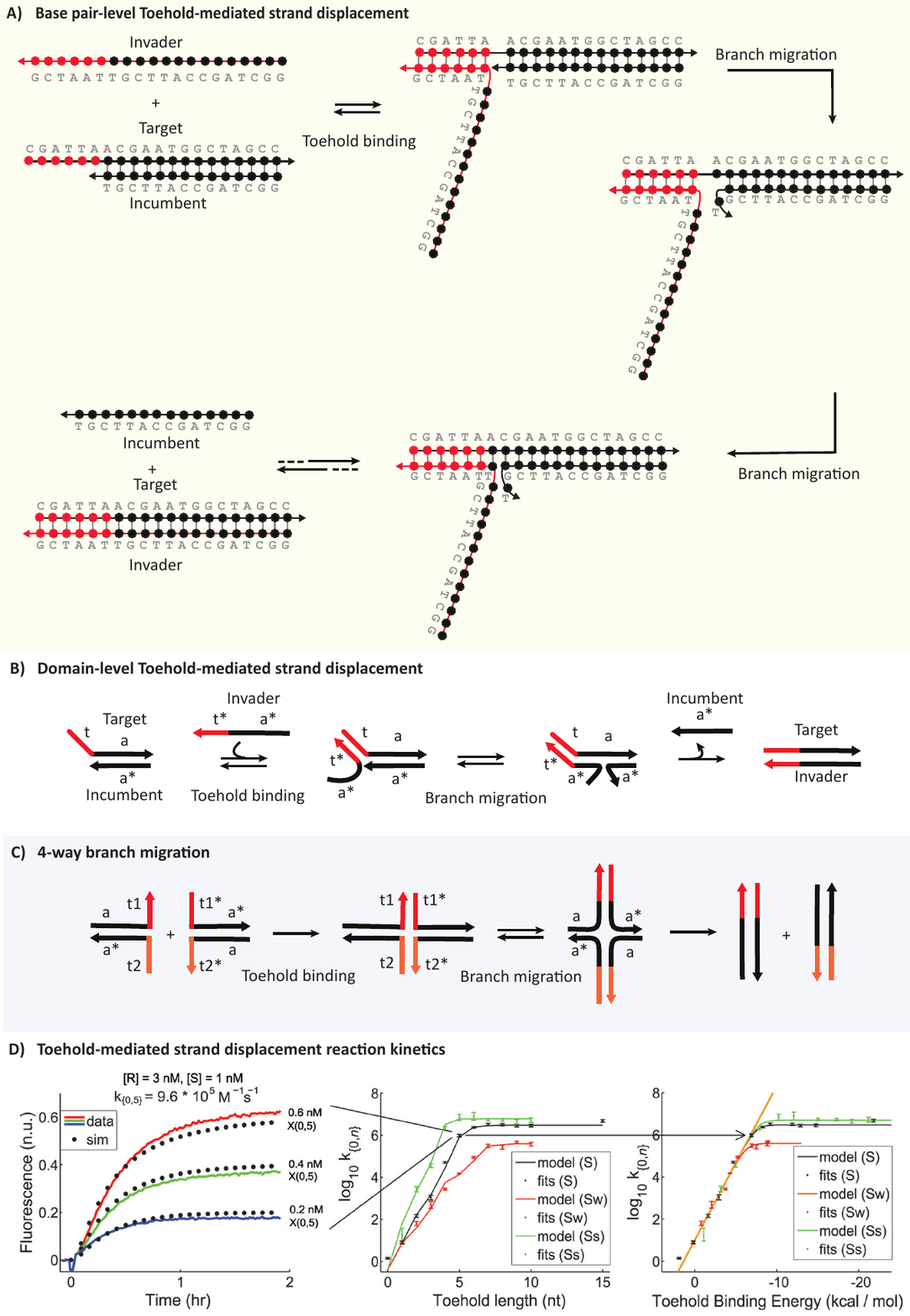}
    \caption{\textbf{Fundamental strand displacement mechanisms and reaction kinetics profiles.}
    \textbf{A–B)} Toehold-mediated strand displacement (TMSD).
    Panel A shows the detailed mechanism at base-pair resolution, with complementary bases represented as black and red dots.
    An invader strand carrying a complementary domain ($t^*$) binds to the exposed toehold of a target–incumbent duplex.
    At the branch point of the resulting three-stranded complex, incumbent and invader compete for base pairing.
    Stepwise branch migration ensues, ultimately displacing the incumbent and yielding a stable target–invader duplex.
    Panel B shows the corresponding primitive, a simplified schematic of the same mechanism.
    \textbf{C)} Four-way branch migration, in which two DNA duplexes hybridise via complementary toehold domains to form a transient Holliday junction.
    Branch migration proceeds through coordinated base-pair exchange at the junction until strand exchange is complete or the system reverts.
    This mechanism is slower than three-way strand displacement but offers higher specificity, greater orthogonality, and avoids single-stranded intermediates.
    \textbf{D)} Kinetic characterisation of TMSD, adapted from Zhang and Winfree~\cite{zhang_control_2009}.
    (i) Fluorescence trajectories showing product formation over time at different invader concentrations for a 5-nt toehold.
    (ii) Logarithm of the fitted rate constants as a function of toehold length, showing an approximately exponential increase before reaching a plateau around 7-nt.
    (iii) The same rate constants plotted against predicted toehold binding free energy, illustrating an initial exponential dependence followed by saturation.
    Thick arrows represent oligonucleotides; complementary domains are colour-coded and labelled with asterisks.}
    \label{fig:overview}
\end{figure}

The programmability of DNA, rooted in sequence-specific base-pairing, has enabled its use as a material for both static and dynamic nanoscale systems. Early work focused on self-assembling architectures such as junction lattices and DNA origami~\cite{seeman_nucleic_1982, rothemund_folding_2006}, while more recent efforts have increasingly exploited control over dynamic processes \cite{zhang_dynamic_2019}. DNA strand displacement (SD) reactions have become the central mechanism for programming dynamic behaviour, following the seminal DNA-fuelled molecular switch of Yurke et al \cite{yurke_dna-fuelled_2000}. Since then, SD has underpinned a wide range of reconfigurable, signal-responsive, and computational architectures \cite{simmel_principles_2019, lv_biocomputing_2021}. Figure \ref{fig:overview}A and~\ref{fig:overview}B show the simplest and most widely used mechanism called toehold-mediated strand displacement (TMSD). A duplex consisting of two complementary strands, termed the target and the incumbent, presents a short unpaired single-stranded "domain" known as the toehold. In the presence of an invader strand with a complementary toehold domain, a strand displacement reaction is initiated that releases the incumbent and forms a stable invader–substrate duplex.

By tuning toehold accessibility, sequence composition, and base-pairing topology, researchers have devised a repertoire of SD variants - including toehold exchange, remote, internal, and associative toeholds, and handhold-mediated displacement - that support conditional logic, cascaded reaction networks, and templated copying. While such mechanisms can be embedded in larger network architectures (for overviews, see e.g.,\ \cite{seeman_dna_2017, chen_dna_2023}), this review focuses on single-reaction mechanisms. We first summarise the key design parameters that govern reaction behaviour and then assess current modelling approaches and their limitations. Finally, we classify SD topologies by recognition-domain geometry, discuss how these factors influence kinetics and function, and outline priorities for predictive, scalable design.

\section{Mechanism, kinetics, and design rationale}

\subsection{Mechanistic Basis of Reaction Kinetics}

Strand displacement begins when the recognition domain of an invader hybridises to an exposed toehold on a target complex. In this three-strand complex the invader and incumbent strand compete for complementary binding to the substrate strand in a stochastic random walk process termed branch migration (Figure \ref{fig:overview}A) \cite{reynaldo_kinetics_2000}. Because toehold binding is transient, efficient displacement of the incumbent strand requires that branch migration outcompetes toehold dissociation. The kinetics are therefore highly sensitive to the toehold design and can be tuned to proceed as fast as possible or to stay within a specific target range: each additional nucleotide can increase the second-order rate constant by approximately one order of magnitude, with saturation near \(10^{7}\,\mathrm{M}^{-1}\,\mathrm{s}^{-1}\) for toeholds of 7–8 nucleotides (nt) (Figure \ref{fig:overview}D) \cite{zhang_control_2009}. Sequence composition further modulates effective rates, with GC-rich toeholds producing longer-lived complexes and thereby increasing the likelihood that branch migration is completed.

The branch migration reaction, unlike the toehold, usually contributes relatively little to the overall reaction rate because it proceeds through base-pair exchange without net base-pair formation \cite{srinivas_biophysics_2013}. Introducing base-pair mismatches can impose position-dependent kinetic barriers that render branch migration rate-limiting. A mismatch proximal to the toehold hinders early migration steps and slows displacement, whereas repairing it removes this barrier and greatly accelerates displacement for short toeholds. In contrast, repairing a distal mismatch has minimal kinetic impact but stabilises the final duplex, introducing a "hidden thermodynamic drive" \cite{machinek_programmable_2014,haley_design_2020}. Mismatches can also initiate displacement by locally destabilising the duplex, enabling toehold-free branch migration \cite{talbot_mismatch-induced_2025}. Although branch migration is generally a minor kinetic factor in DNA-only systems, DNA-RNA hybrids form a clear exception: purine-rich sequences and base distribution strongly affect migration rates, allowing tuning over four orders of magnitude through sequence design \cite{smith_strong_2023,ratajczyk_controlling_2025}.

These principles dictate the biophysics of TMSD in its simplest form (Figure \ref{fig:overview}D). Alternative "topologies" - referring to the arrangement of base-pairing domains - allow distinct logical behaviour and strongly influence reaction kinetics (Table \ref{table:relative_sd_topologies}). We will first examine the sequence and structural constraints that shape - and often limit - the design space before considering the range of SD topologies in section~\ref{domain_topology}.

In its simplest form, a TMSD reaction can be designed by hand using Watson–Crick base-pairing rules~\cite{watson_molecular_1953}. Functional systems, however, often involve multiple interacting strands and conditional pathways. Strands must encode all intra-strand structures (e.g., hairpins) and inter-strand assemblies (duplexes, junctions), that are required at any stage in the system's operation. Moreover, unintended interactions and misfolding must be designed out~\cite{jr_thermodynamics_2004}. As complexity increases, the space of valid sequences narrows, making effective design increasingly difficult.

\begin{figure}[H]
    \centering
    \includegraphics[width=0.45\textwidth]{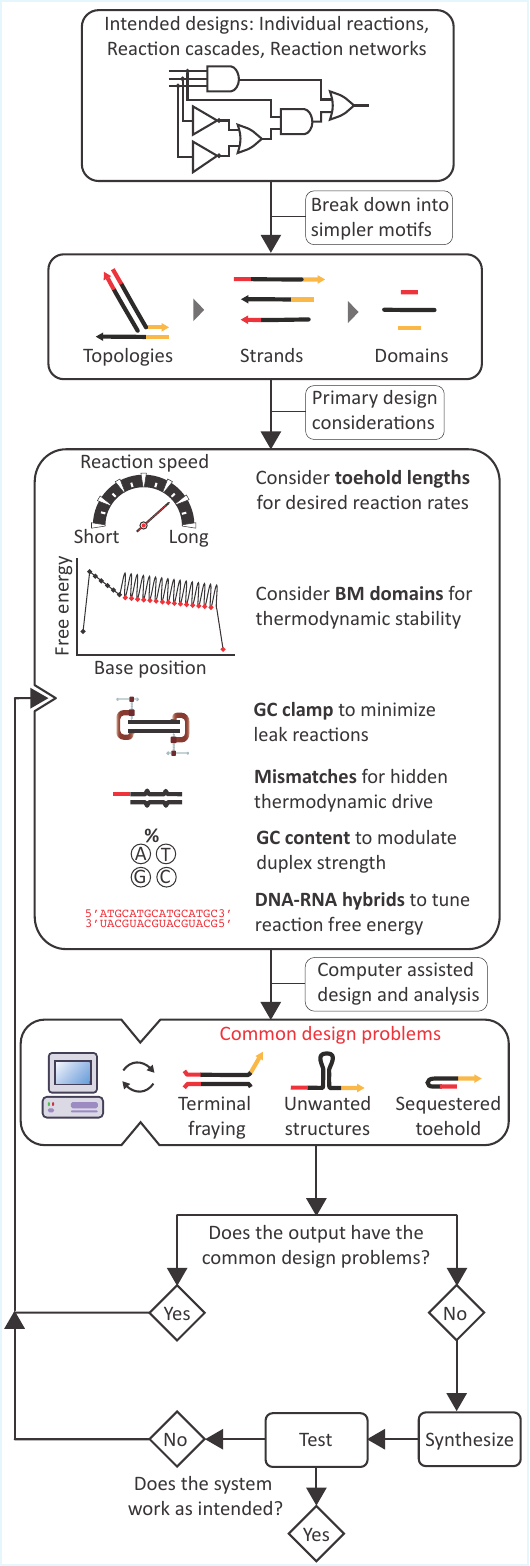}
    \caption{\textbf{Computational-experimental workflow for strand displacement system design.} An iterative design approach begins by dividing fundamental parameters of the target architecture into manageable motifs. Primary design choices - such as toehold length and branch migration (BM) stability - define constraints that narrow the design space. Computational tools then analyse candidate sequences within this space, followed by experimental validation of the most promising designs. Through iterative refinement, the system is checked for common issues (terminal fraying, toehold sequestration, unwanted secondary structures) until it achieves the desired performance.}
    \label{fig:workflow}
\end{figure}

Automated frameworks, such as peppercorn, the domain-level reaction enumerator by Badelt et al. \cite{badelt_domain-level_2020}, help generate sequences of DNA strands starting from a high-level chemical reaction networks. However, for novel topologies or unconventional pathways, an iterative, experiment-guided design approach remains essential. Figure \ref{fig:workflow} illustrates one methodology for approaching the design process. The broader challenge is that a fully integrated, general-purpose design pipeline for SD networks is yet to be demonstrated across diverse topologies and reaction mechanisms.

\subsection{Leak Reactions and Mitigation}

Even well-designed systems can exhibit leak: undesired strand displacement or hybridisation arising from duplex fraying, out-of-register or partial base pairing, or other spurious interactions. Several strategies can suppress leak \cite{wang_effective_2018}: terminal G–C pairs reduce fraying, and clamps - base pairs present in an initial duplex but absent in a product - block toehold-free invasion. Toehold sequestration can also be used, burying the recognition domain in a hairpin until its triggered \cite{yang_regulation_2016,green_dna_2006}. Wang et al. combined these insights into a systematic redundancy-based design, requiring multiple long domains to associate and suppressing leak below detection even at high concentrations~\cite{wang_effective_2018}. Derauf and Thachuk instead proposed polymerase-dependent TMSD, where enzymatic extension releases output only after a successful toehold exchange, thereby maintaining fidelity without the kinetic penalties of redundancy \cite{derauf_leakless_2025}. Together, these approaches suppress leak by thermodynamic penalties or kinetic checkpoints but leave residual leak and typically add complexity to the SD system.

\subsection{Modelling tools and predictive limitations}
\label{Modelling tools and predictive limitations}

Computational tools are central to the rational design of SD systems. NUPACK \cite{wolfe_constrained_2017} is widely regarded as the main design tool, supporting sequence design under structural and thermodynamic constraints. Such tools are fast enough for practical design because they rely on equilibrium-state predictions rather than explicit kinetics. By contrast, coarse-grained molecular dynamics frameworks such as oxDNA \cite{machinek_programmable_2014} and its RNA variant oxRNA/oxNA \cite{ratajczyk_coarse-grained_2024} provide detailed insight into strand motion, conformational transitions, and reaction pathways, but are computationally too demanding for sequence-level design. Base-pair level kinetic models, such as Multistrand \cite{schaeffer_stochastic_2015} and related approaches \cite{irmisch_modeling_2020}, attempt to bridge this gap and can be parametrised to match specific processes (e.g., conventional TMSD) but have so far seen limited generality across more complex topologies. Machine learning approaches have begun to emerge but are not yet in common use. These models aim to extract predictive features directly from sequence data~\cite{wang_predicting_2024, akay_predicting_2024, long_understanding_2024}, but remain limited by small training datasets and the inherent complexity of strand-exchange dynamics.

The challenges of rate prediction were recently highlighted in the review by Ashwood and Tokmakoff: hybridisation occurs on a rugged energy landscape populated by transient, often non-productive encounter complexes~\cite{ashwood_kinetics_2025}. Features such as out-of-register base pairing, metastable folding, and ionic conditions strongly influence reaction rates, yet remain poorly characterised in most models. Thus, while thermodynamic structure prediction using nearest-neighbour models is robust and broadly reliable, quantitative prediction of reaction kinetics remains an open problem. Consequently, current design strategies rely on computational simulation followed by systematic empirical refinement, as exemplified by the workflow in Figure \ref{fig:workflow}.

\begin{figure}[H]
    \centering
    \includegraphics[width=\textwidth]{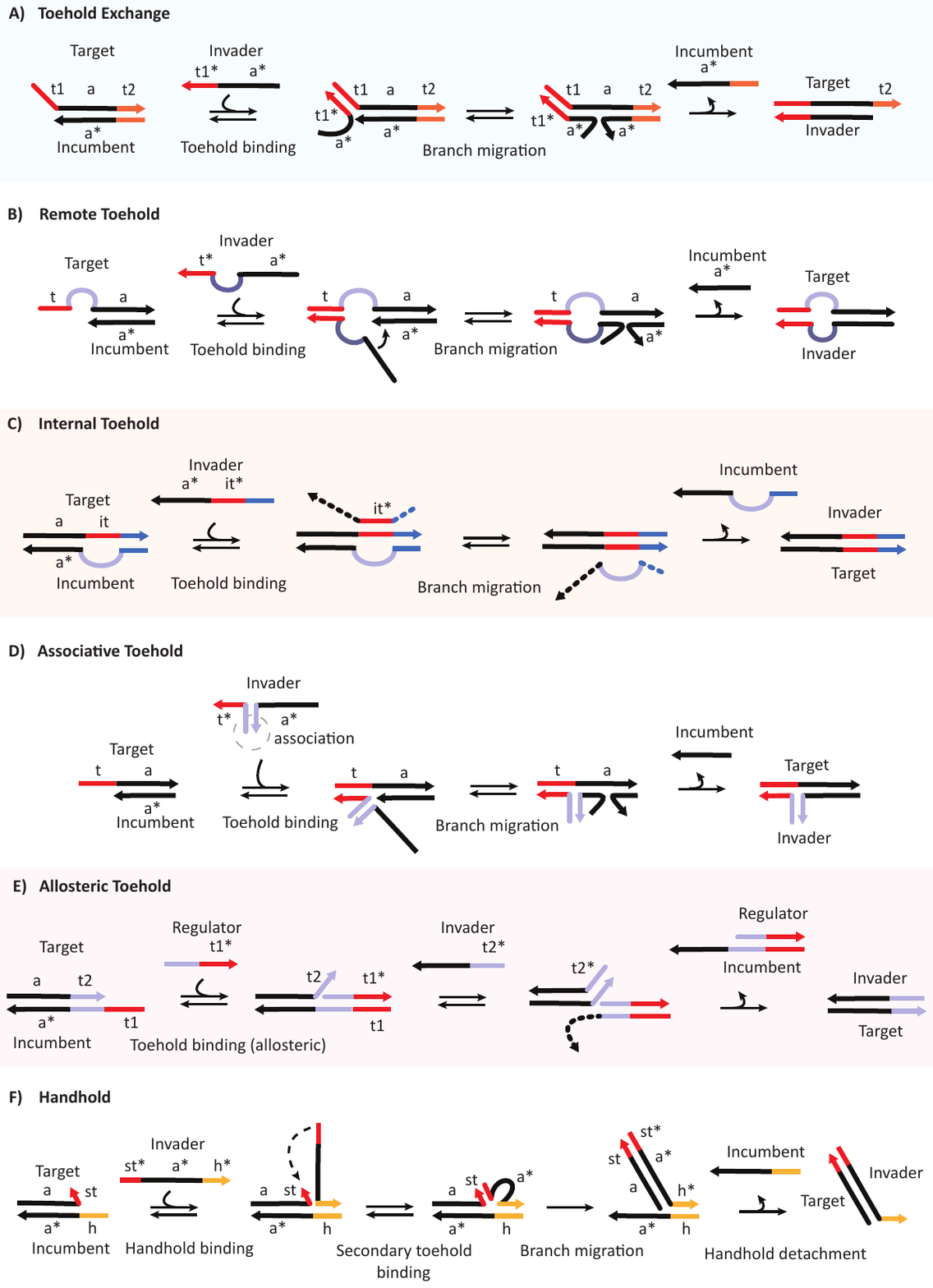}
    \caption{\textbf{Strand displacement topologies.} Schematic representations of individual mechanisms. Thick arrows denote oligonucleotides; complementary domains are colour-coded and labelled with asterisks. Abbreviations: t, t$_1$, t$_2$ = toeholds; st = secondary toehold; it = internal toehold; h = handhold; a = branch migration domain.}
    \label{fig:TOPOLOGIES}
\end{figure}

\subsection{Strand Displacement Topologies}
\label{domain_topology}

\begin{table}[b!]
\centering
\caption{Relative second-order rate constants ($k_{1}$) of strand-displacement topologies, compared against classic TMSD under standard conditions (25~$^\circ$C, 1$\times$ TAE, 1~M NaCl or 12.5~mM Mg$^{2+}$).}
\label{table:relative_sd_topologies}
\begin{tabularx}{\textwidth}{lXl}
\toprule
\textbf{Topology} & \textbf{Relative to classic TMSD} & \textbf{Reference} \\
\midrule
Classic TMSD & Reference standard. Tunable over $k_{1}$ up to $\sim 10^{7}$ M$^{-1}$\,s$^{-1}$ depending on toehold length and GC content. & \cite{zhang_control_2009,srinivas_biophysics_2013} \\
\addlinespace[3pt]
Toehold-exchange & Comparable reaction speed if the output toehold is $\leq$6 nucleotides. Enables reversible cascades. & \cite{zhang_control_2009,zhang_dynamic_2011} \\
\addlinespace[3pt]
Remote toehold & Typically 10- to 1000-fold slower. A spacer separates the toehold from the branch point; longer and more flexible spacers reduce the effective concentration. & \cite{genot_remote_2011} \\
\addlinespace[3pt]
Internal toehold & Typically 10- to 100-fold slower. Branch migration initiates from a loop; constrained geometries impose an entropic barrier. & \cite{green_dna_2006} \\
\addlinespace[3pt]
Associative toehold & Slower to comparable. Without bulges, typically 10- to 100-fold slower; suitable bulges/coaxial stacking can restore near-TMSD rates. & \cite{chen_expanding_2012} \\
\addlinespace[3pt]
Allosteric toehold & Typically 10- to 1000-fold slower. Strand gating reduces access to the recognition domain. & \cite{yang_regulation_2016,c_programmable_2022} \\
\addlinespace[3pt]
Handhold-mediated & With equal toehold and handhold lengths, typically slower. Transient templating via the incumbent strand; rate is highly design-dependent. & \cite{cabello-garcia_handhold-mediated_2021,mukherjee_kinetic_2024} \\
\addlinespace[3pt]
4-way SD & Branch migration is intrinsically slower; limiting rates at high concentration of strands remain below TMSD, and short toeholds are very slow. & \cite{bakhtawar_overcoming_2025,dabby_synthetic_2013} \\
\addlinespace[3pt]
RNA-DNA hybrids & Reaction rates are highly sensitive to the branch-migration sequence. Fastest reactions consistent with DNA-DNA speeds. & \cite{smith_strong_2023,ratajczyk_controlling_2025} \\
\bottomrule
\end{tabularx}

\smallskip
\footnotesize
\textbf{Environmental factors.} Rate constants assume standard buffer conditions. Increasing temperature generally accelerates both toehold binding and branch migration. Mg$^{2+}$ stabilises junctions but can suppress certain topologies (e.g., four-way SD or DNA$\rightarrow$RNA) \cite{zhang_dynamic_2011,srinivas_biophysics_2013,ratajczyk_controlling_2025}.
\end{table}

\subsubsection{Toehold-Exchange Reactions}
Figure \ref{fig:TOPOLOGIES}A shows a toehold exchange reaction. The invading strand initiates branch migration through a toehold, but a few terminal base pairs at the end of the displacement domain are deliberately excluded from branch migration. These remaining pairs spontaneously dissociate, exposing a new domain on the substrate strand that can serve as an output toehold (thus "exchange" of toeholds). This mechanism expands the logical space of SD, enabling logic gates in which molecules act as both products and inputs \cite{zhang_control_2009}, with notable implementations in catalytic systems based on paired-junction structures \cite{wang_design_2024}.

\subsubsection{Remote Toehold Reactions}
Remote-toehold reactions insert a flexible spacer between the toehold and the branch-migration domain, decoupling binding from branch migration and providing a kinetic handle (Fig. \ref{fig:TOPOLOGIES}B). Genot et al. varied a poly-T linker from 1~nt to 23~nt, modulating reaction rates by more than three orders of magnitude without altering the branch-migration or toehold sequence. Converting the spacer to double-stranded DNA or polyethylene glycol provides further control. This versatility supports toehold reuse, leak suppression in autocatalytic loops, precise rate tuning or thresholding in multilayer cascades, and discrimination of single-base mismatched DNA \cite{genot_remote_2011, li_modulating_2016}.

\subsubsection{Internal Toehold Reactions}
In the internal toehold topology, a short single‑stranded domain is buried in a bulge or hairpin; an invader binds this internal site and then drives branch migration from the inside out (Figure \ref{fig:TOPOLOGIES}C). Green et al. \cite{green_dna_2006} found that opening from an internal toehold is 10–100 times slower than from an equivalent external toehold positioned at the free end of the strand. Despite this fact, internal toeholds of 5–7 nucleotides still enable complete opening of the hairpin stem when the stem is at least 16 base pairs long. Green et al. also exploited complementary hairpin motifs as chemical fuel to power catalytic reactions.

\subsubsection{Associative Toehold Reactions}

Associative toehold activation links the toehold and branch migration domains through hybridisation of auxiliary strands. This strategy allows strand displacement networks to be assembled in a combinatorial fashion at runtime rather than predefined during synthesis (Fig. \ref{fig:TOPOLOGIES}D). Chen et al. \cite{chen_expanding_2012} demonstrated that a three-way junction stabilised by two bulged thymidines enables efficient strand displacement, where enhanced coaxial stacking lowers the activation barrier and allows dynamic reassignment of toehold–branch migration pairings (“toehold switching”) to expand circuit architectures.

\subsubsection{Allosteric Toehold}

The allosteric toehold is a modular strand displacement topology in which the toehold and branch migration domains of the invader are split across two strands: a regulator and an input (Fig. \ref{fig:TOPOLOGIES}E). The regulator displaces a short domain in a target–incumbent duplex, exposing a toehold for the invader. This mechanism enables reversible and conditional activation, supporting programmable gating, selective triggering, and flexible reuse of components in logic circuits, catalytic amplifiers, and reconfigurable DNA devices \cite{yang_regulation_2016}. Additionally, Weng et al. developed a hybrid mechanism termed cooperative branch migration, where displacement occurs only when two independent inputs - a distal toehold and an allosteric regulator - act together to reconfigure the recognition domain~\cite{weng_cooperative_2022}.

\subsubsection{Handhold-Mediated Displacement}

Handhold-mediated strand displacement (HMSD) \cite{cabello-garcia_handhold-mediated_2021} repositions the recognition domain onto the incumbent strand, creating a transient binding site that promotes initial invader–target contact without leaving a permanently bound recognition domain (Fig. \ref{fig:TOPOLOGIES}F). Cabello-García et al. showed that the topology of HMSD enables an incumbent strand to act as a catalytic template, accelerating the association of two other strands, and Mukherjee et al. later demonstrated how this mechanism can be harnessed for kinetic proofreading \cite{mukherjee_kinetic_2024}.

\subsubsection{Four-way branch migration reactions}
In four-way branch migration, a Holliday junction is formed when complementary toeholds on two duplexes hybridise and initiate a strand exchange (Fig.~\ref{fig:overview}C) \cite{dabby_synthetic_2013}. The absolute speed limit is lower than conventional TMSD because each migration step requires the concerted exchange of two base pairs and is stabilised by junction stacking. Bakhtawar et al.\ showed that single-nucleotide bulges destabilise junction stacking, enable half-step migrations, and accelerate branch migration over tenfold while increasing junction unbinding \cite{bakhtawar_overcoming_2025}. Four-way branch migration thus offers a complementary topology: while intrinsically slower, it avoids free single-stranded reactants and enhances specificity through dual-toehold recognition, which can be advantageous for applications in vivo or complex environments \cite{simmel_principles_2019}.

\subsubsection{Other Kinetic Modulation Strategies}

Beyond canonical SD topologies, various auxiliary strategies modulate reaction kinetics through structural, enzymatic, and chemical means. The following examples accelerate reaction rates: Singh et al.\ showed that the ring-shaped Twinkle helicase catalyses SD by locally unwinding base pairs, increasing rates up to 6000-fold over the uncatalysed case \cite{singh_twinkle-catalyzed_2023}. Zhang et al.\ introduced irreversible toehold binding via covalent click chemistry, achieving a $\sim$1000-fold acceleration for short toeholds \cite{zhang_click_2025}. Liu and Zhang reported that the cationic polymer polyquaternium-2 (PQ2) enhances TMSD by condensing DNA strands and increasing effective concentration; even 1-nt toeholds exhibited a $10^5$-fold rate increase in the presence of PQ2 \cite{liu_accelerating_2025}.

There are also several strategies to slow or suppress reaction rates: Bai et al. used a tweezer motif for signal transduction and modulation of reaction kinetics \cite{bai_spatially_2024}, while Lysne et~al. introduced steric barriers using star-shaped motifs to reduce invader access to recognition domains \cite{lysne_leveraging_2023}. An additional reversible suppression strategy is the antitoehold (At) \cite{wu_plug-and-play_2024}: a short strand that binds the exposed toehold and forms a metastable clamp, thereby lowering the effective on‑rate of TMSD without significantly altering the reaction’s thermodynamic end state. Addition of a complementary anti‑At releases the clamp and restores the toehold, enabling tunable TMSD rates over large dynamic ranges and with on/off control.

Temporal control can also be achieved using dissipative systems. In such systems, continuous energy consumption is required to maintain the active state, meaning that the reaction dynamics are driven away from thermodynamic equilibrium and are inherently transient. Bucci et al. engineered enzyme-regulated pulse generators that couple fuel consumption to strand displacement, producing reversible activation–deactivation cycles \cite{bucci_timed_2023}. These cycles generate pulses of SD activity that decay once the fuel is exhausted, thereby enabling time-gated signalling without altering the intrinsic rate constants of the displacement reaction.

Finally, Nikitin introduced a mechanistically distinct paradigm termed strand commutation \cite{nikitin_non-complementary_2023}. Unlike conventional strand displacement, which relies on fully complementary interactions and branch migration, strand commutation exploits partially complementary or even non-complementary strands that reversibly assemble into weak complexes. Instead of a single invader displacing an incumbent, all strands participate in a dynamic ensemble governed by the law of mass action. This framework enables both digital and analogue computation without toeholds, hybridisation cascades, or branch migration, marking a conceptual departure from established SD mechanisms.

\section{Outlook}

Strand displacement systems are versatile: from first principles they support a diverse repertoire of design primitives that can be composed into increasingly complex circuits. Current approaches allow empirical tuning of reaction kinetics through toehold and branch migration design, mismatches, clamps, and backbone composition, but their outcomes cannot be faithfully predicted \textit{a priori}.

This limitation highlights a broader challenge: the need for an integrated framework that links mechanistic understanding to scalable system design. For standardised, large-scale dynamic design we need a route from a symbolic reaction specification to concrete sequences and topologies with defined kinetic and leakage properties. A useful target is a logic syntax that encodes networks and desired rates/leaks, e.g.,\ $\mathrm{AX}+\mathrm{B}\;\rightleftharpoons\;\mathrm{AB}+\mathrm{X}$; $\mathrm{AB}+\mathrm{Y}\;\rightleftharpoons\;\mathrm{BY}+\mathrm{A}$, annotated with bounds such as reaction rate windows, maximum leak, temperature, and ionic conditions. A “compiler” could then propose candidate designs from the repertoire of mechanisms reviewed here, but present approaches still provide limited fidelity for predicted absolute reaction rates and limited control over leak. Realising such a compiler will require broad, standardised kinetic datasets reporting $k_{\mathrm{on}}$, $k_{\mathrm{off}}$, branch-migration rates, and leak with uncertainties across sequence families and buffers, alongside hybrid molecular-dynamics and machine-learning models for quantitative prediction. Meeting these needs is key to scalable, predictive design of dynamic DNA systems. A further challenge is translation to \textit{in vivo} settings, where stability, delivery, and molecular crowding must be addressed \cite{jung_test_2025}.

\newpage
\section*{Funding}

This work was supported by a Royal Society University Research Fellowship Renewal and associated Expenses (grant no. URF\textbackslash R\textbackslash 211020 to T. E. O., RF\textbackslash ERE\textbackslash 231045 to M. M., and RF\textbackslash ERE\textbackslash 210246 to K. J.). This work is part of a project that has received funding from the European Research Council (ERC) under the European Union's Horizon 2020 research and innovation programme (grant agreement 851910 to M. M., R. M., and T. E. O.).

\section*{Declaration of Interest}

The authors declare no conflict of interest.

Portions of this manuscript were edited and language-checked using a large language model (ChatGPT, OpenAI). The authors reviewed and approved all content prior to submission.

\section*{Credit Author Statement}
Križan Jurinović: Conceptualisation, Writing -- original draft, Visualisation, Writing -- review \& editing.\\
Thomas E. Ouldridge: Conceptualisation, Supervision, Writing -- review \& editing.\\
Merry Mitra: Visualisation, Writing -- review \& editing, Conceptualisation.\\
Rakesh Mukherjee: Visualisation, Writing -- review \& editing, Conceptualisation.

\newpage
\section*{References and recommended reading}
Papers of particular interest, published within the period of review, have been highlighted as:

* of special interest \newline
** of outstanding interest

\printbibliography

\end{document}